\documentclass[aps,prl,twocolumn,floatfix,%
superscriptaddress,showpacs]{revtex4}
\usepackage{graphicx}

\begin{document}

\title{Polarized excitons in nanorings and the `optical'
Aharonov-Bohm effect}

\author{A. O. Govorov}
\affiliation{Department of Physics and Astronomy, and Nanoscale
and Quantum Phenomena Institute, Ohio University, Athens, Ohio
45701-2979}
 \affiliation{Institute of Semiconductor Physics, Russian Academy
of Sciences, Siberian Branch, 630090 Novosibirsk, Russia}

\author{S. E. Ulloa}
\affiliation{Department of Physics and Astronomy, and Nanoscale
and Quantum Phenomena Institute, Ohio University, Athens, Ohio
45701-2979}

\author{K. Karrai}
\affiliation{Center for NanoScience, LMU, Geschwister-Scholl-Platz
1, 80539 M\"unchen, Germany}

\author{R. J. Warburton}
\affiliation{Department of Physics, Heriot-Watt University,
Edinburgh, UK}

\date{\today } 

\begin{abstract}
The quantum nature of matter lies in the wave function phases that
accumulate while particles move along their trajectories. A
prominent example is the Aharonov-Bohm phase, which has been
studied in connection with the conductance of nanostructures.
However, optical response in solids is determined by neutral
excitations, for which no sensitivity to magnetic flux would be
expected.  We propose a new mechanism for the topological phase of
a neutral particle, a polarized exciton confined to a
semiconductor quantum ring.  We predict that this magnetic-field
induced phase may strongly affect excitons in a system with
cylindrical symmetry, resulting in switching between `bright'
exciton ground states and novel `dark' states with nearly infinite
lifetimes. Since excitons determine the optical response of
semiconductors, the predicted phase can be used to tailor photon
emission from quantum nanostructures.
\end{abstract}

\pacs{78.67.Hc, 73.21.La, 73.23.Ra}

\maketitle

It is known that the quantum mechanical phase of a state wave
function is not a physical observable.  This understanding, true
in its absolute form, does not preclude the important possibility
of observing \textit{relative phases} in a suitably prepared
system.  In fact, much of the physics in meso- and nanoscopic
systems is intrinsically connected to interference or phase-shift
phenomena that manifest themselves in a number of experimentally
measurable quantities.  Prominent examples include the
superconducting quantum interference devices (SQUIDs) \cite{1} the
quantum-phase factors induced by adiabatic changes (known as
geometric Berry phases) \cite{2}, their generalization to
non-adiabatic changes due to Aharonov and Anandan \cite{3}, and
the well known Aharonov-Bohm (AB) effect \cite{4}.  Such a case
appears naturally in systems with ring geometry in the presence of
a magnetic field. In fact, recently available semiconductor rings
$\simeq 10$--100nm in diameter allow one to explore this
interesting physics in readily attainable magnetic fields. We
report here on a new mechanism of phase difference acquired in a
magnetic field by a \textit{composite} and {\em polarizable}
object with overall \textit{zero charge}. Such neutral particles,
excitons, are bound states of an electron and a hole in
semiconductors, and are responsible for optical emission of
crystals at low temperatures. The predicted interference effect
has important observable consequences, as it affects the exciton
emission and lifetime in nanoscopic semiconducting rings, and
provides a novel phase interference phenomenon, the `optical' AB
effect.  In particular, we predict a striking effect: the exciton
emission can be strongly suppressed in certain magnetic-field
windows.

The AB phase is most simply described as the phase acquired by a
charge as it traverses a region where a magnetic flux exists,
\textit{while no effects of the classical Lorentz force are
present}.  The AB effect has been verified in a number of
beautiful experiments using superconducting rings \cite{5}, where
electrons move in the regions with zero magnetic field.  In
semiconductors, the AB effect has been used in fascinating
experiments to measure the relative phases of alternative paths in
electronic interferometers \cite{6}, and proposed as a mechanism
to distinguish singlet and triplet states in coupled quantum dots,
in connection with their use in solid state quantum computers
\cite{7,8}.

The exciton bound state of electron and a corresponding hole is a
neutral object basically insensitive to external magnetic fields
or fluxes (except for a weak diamagnetic behavior).  In fact,
excitons in quantum dots a few tens of nanometers in radius and an
even smaller height, have been studied in recent times
\cite{9,10,11,12}. Moreover, thanks to the infrared absorption
studies of Lorke \textit{et al}., we know that semiconductor
quantum rings are possible \cite{13}, where the electrons are
coherent around the ring and non-localized. In this type of
structures excitons in \textit{individual} nanorings in strong
magnetic fields can be generated and monitored using near-field
scanning optics techniques \cite{9,14}. Indeed,
micro-photoluminescence ($\mu$PL) experiments in individual
quantum rings would provide an opportunity to explore the AB
effect in neutral excitons \cite{15,16}.  We should emphasize that
the AB oscillations of the ground state are predicted to be {\em
negligible} at typical size/fields in the 1D rings \cite{16}.
However, as we explain below, the in-plane radial polarization of
the exciton in finite-width quantum rings produces a strong
modulation of the oscillator strength of the excitonic transition
with magnetic field, due to the AB effect.

Figure 1a illustrates Berry's gedanken experiment to explain how
the adiabatic evolution of the Hamiltonian in a system gives rise
to an additional phase \cite{2}.  The Berry phase is accumulated
as the particle placed in an isolating box is slowly rotated
around the field lines.  Since the evolution is adiabatic, the
system returns to its original state after a full cycle, and only
a topological phase accumulates.  The phase is given by the total
flux in terms of the flux quantum, $\Phi/\Phi_0$, and is
equivalent to the AB phase.  When the object is a composite of two
particles, one can imagine a similar adiabatic evolution, although
here we rotate a \textit{polarized} exciton having a radial dipole
moment. After one full revolution, the particles obtain
topological phases again, but these phases will be
\textit{different}, since the electron and hole have different
trajectories.  The phase difference is due to the unequal magnetic
fluxes penetrating their paths. It is important, however, that the
net relative phase would be effectively sampled/measured when the
two carriers in the system combine to emit the photon seen in the
PL experiment.  The accumulated phase shift depends intrinsically
on the polarizability of the composite object, and can even make
optical transitions impossible, resulting in suppression of
optical emission in well-defined windows of magnetic field. We
demonstrate this effect using a model of a polarized exciton
rotating in a quantum ring (QR). Notice that topological phases on
point dipoles have been considered \cite{17}, but their internal
structure was not of interest in that case.

Today's semiconductor technology offers a variety of
nanometer-size structures, including self-assembled QR's
\cite{13,18}, and type-II quantum dots \cite{19,20}.  In a
ring-shaped nanostructure, any asymmetry in the confinement
potential may cause the polarization of excitons in the radial
direction (Fig.\ 1c), so that an electron and hole may move along
different circles. In fact, the existence of such an asymmetry is
natural since an electron and hole have very different masses and
move in different potentials, $U_e$  and $U_h$ \cite{21}. Namely,
in self-assembled QR's (Fig.\ 1c), the lighter electron tends to
tunnel toward the ring center, whereas the heavier hole is
localized mostly near the ring radius, producing a net radial
polarization \cite{20}. Such polarization is easily calculated
from a parameterized model of a quantum ring, $U_{e(h)} =m_{e(h)}
\Omega^2_{e(h)}(r_{e(h)}-R_r)^2/2$, where $r$ is the distance to
the center of the ring, and $m_{e(h)}$ and $\Omega_{e(h)}$ are the
effective masses and single-particle confinement frequencies,
respectively.  Moreover, the radial polarization in the exciton
can appear also due to the deformation potential in a QR, which
can produce drastically different effective potentials in the
valence and conduction bands.  The structural in-plane radius of a
QR determined from atomic force micrographs is about 40nm (Fig.\
1c), however, the electronic radius $R_r$ estimated from optical
experiments turns out to be much smaller, $R_r \simeq 16$nm
\cite{13}.  As the vertical, $z$-extension of the QR's is much
smaller than their lateral size, we consider here only the
in-plane motion, assuming strong quantization in the
$z$-direction. In the case of type-II quantum dots, the radial
dipole moment in an exciton appears because the quantum dot
structure potential in its center forms a barrier for electrons
and a well for holes. This situation results then in the electron
and hole moving along different circles \cite{22}. To describe
excitons in both of these systems, we use a model of two
`parallel' one-dimensional rings, one for the electron and another
for the hole (insert in Fig.\ 2), having different radii, $R_e$
and $R_h$, respectively.

The Hamiltonian of the exciton in this model reads
 \begin{equation}
\hat{H}_{tot}=\hat{T}_e+\hat{T}_h+U_{e}+U_{h}+ U_{C}(|{\bf
r}_e-{\bf r}_h|), \label{H_tot}
 \end{equation}
where ${\bf r}_{e(h)}$ are the in-plane coordinates,
$\hat{T}_{e(h)}$ are the kinetic energies in the presence of a
normal magnetic field, and $U_{C}$ is the Coulomb potential. Now
we assume that the quantization in the radial direction is
stronger than that in the azimuthal direction. It allows us to
separate variables in the exciton wave function, $\Psi({\bf
r}_e,{\bf r}_h)=f_e(\rho_e)f_h(\rho_h)\psi(\theta_e,\theta_h)$.
Here ${\bf r}=(\rho,\theta)$. The radial wave functions $f_{e(h)}$
are strongly localized near the radii, $R_{e(h)}$. The Hamiltonian
describing the angular wave function $\psi(\theta_e,\theta_h)$  is
(up to a $B$-independent constant term),
\begin{eqnarray}
 \hat{H}_{exc} & =&
 - \frac{\hbar^2}{2m_eR_e^2}\frac{\partial^2}{\partial\theta_e^2}
 - \frac{i\hbar\omega_e}{2}\frac{\partial}{\partial\theta_e}
 - \frac{\hbar^2}{2m_hR_h^2}\frac{\partial^2}{\partial\theta_h^2}
 \nonumber \\ &+&
 \frac{i\hbar\omega_h}{2}\frac{\partial}{\partial\theta_h}
 + \frac{m_e\omega_e^2R_e^2+m_h\omega_h^2R_h^2}{8}
 \nonumber \\  &+& u_{C}(|\theta_e-\theta_h|) \, ,
 \label{h}
\end{eqnarray}
where $\omega_{e(h)}=|e|B/[m_{e(h)}c]$ are the cyclotron
frequencies of the particles, $B$ is the normal magnetic field,
and $u_c$ is the Coulomb potential averaged over the coordinate
$\rho$ involving the radial wave functions. Now it is convenient
to introduce new variables, $\Delta \theta = \theta_e - \theta_h$,
and $\theta_0=(a \theta_e+ b \theta_h)/(a+b)$, where
$\theta_{e(h)}$ is the angular coordinates of each particle,
$a=m_e R_e^2$, and $b=m_h R^2_h$. The Hamiltonian then reads
$\hat{H}_{exc} = \hat{H}_{rot} (\theta_0) + \hat{H}_{int}(\Delta
\theta)$, where the translation of the whole exciton around the
ring and the internal motion, are separate.  The eigenfunctions
are of the form
$\psi(\theta_e,\theta_h)=\psi_0(\theta_0)\psi_1(\Delta\theta)$.
Here, $\hat{H}_{rot}$ is given by
 \begin{equation}
  \hat{H}_{rot}(\theta_0)= \varepsilon_0 \left[
 -i\frac{\partial}{\partial\theta_0} +\frac{\Phi_{\Delta R
 }}{\Phi_0} \right]^2, \label{Hrot}
 \end{equation}
where  $\varepsilon_0=\hbar^2/(2R_0^2M)$, $R_0=(R_e+R_h)/2$,
$M=(m_eR_e^2+m_hR_e^2)/R_0^2$, and $\Phi_{\Delta
R}=\pi(R_e^2-R_h^2)B=2\pi\Delta R R_0B$ is the magnetic flux
penetrating the area between the electron and hole trajectories
(Fig.~1b); and $\Delta R=R_e-R_h$.

The internal-motion Hamiltonian $\hat{H}_{int}$ involves the
Coulomb potential. In small QR's, the quantization due to kinetic
motion is much stronger than the Coulomb interaction, and the
picture is basically that of single particles. This limit
corresponds to $R_{e(h)}< a_0^*$, where $a_0^*$ is the effective
Bohr radius in the semiconductor. In large QR's, where $R_{e(h)}>
a_0^*$, the motion becomes strongly correlated, as the particles
form a tightly-bound exciton and move together around the ring.

It is possible to obtain analytic solutions for the spectrum in
these two limiting cases.  For small QR's, the electron and hole
move independently (Fig.\ 2, insert). There, the exciton spectrum
is given by
 \begin{equation}
 E_{exc} = E_g + \frac{\hbar^2}{2m_eR_e^2} \left( L_e +
 \frac{\Phi_e}{\Phi_0} \right)^2 + \frac{\hbar^2}{2m_h R_h^2}
 \left( L_h - \frac{\Phi_h}{\Phi_0} \right)^2 \, ,
 \label{no1}
 \end{equation}
 where $E_g$ is the magnetic field-independent term which includes
the band-gap energy. $L_e$ and $L_h$ represent the angular
momenta. The magnetic field $B$ enters Eq.\ (\ref{no1}) through
the magnetic fluxes $\Phi_{e(h)}= \pi R^2_{e(h)}B$, which describe
the quantum phase accumulating in the wave function of each
particle as it travels along the ring.

For the limit of the tightly-bound exciton (Fig.\ 3, insert), the
spectrum is determined by $\hat{H}_{rot}$ and takes the form
 \begin{equation}
  E_{exc}^\prime = E_g^\prime + \frac{\hbar^2}{2 MR_0^2} \left( L +
 \frac{\Delta \Phi}{\Phi_0} \right)^2 \, ,
 \label{no2}
 \end{equation}
where $L= L_e+L_h$ is the total momentum, $M$, $R_0$ are as above,
and $E_g^\prime$ is field-independent.  The net magnetic flux
through the area between the electron and hole trajectories,
$\Delta \Phi$, is proportional to the radial dipole moment of the
exciton, $D=-e(R_e- R_h)$.

The $B$-dependent terms from Eqs.\ (\ref{no1}) and (\ref{no2}) are
shown in the upper panels of  Figs.\ 2 and 3.  Since emission
spectra of semiconductors at low temperatures are governed by the
exciton ground-state, one must pay special attention to the lowest
energy states. The figures show that the ground state changes with
increasing magnetic field, and may acquire non-zero angular
momentum values. For small quantum rings (Fig.\ 2), the ground
state $(L_e,L_h)=(0,0)$ changes successively in favor of the
states (0,+1), ($-1$,+1), ($-1$,+2), etc., as the field increases,
producing a sequence of ground state total angular momentum values
of 0, 1, 0, 1, etc.  On the other hand, for the
strongly-correlated exciton in large rings (Fig.\ 3), the
ground-state momentum changes increasingly from $L=0$ to
$L=+1$,+2,+3 \ldots, with magnetic field.

Only the excitons with zero total momentum can emit photons. This
condition arises as the photon's angular momentum is exhausted by
the atomic wave functions involved in the optical transition. The
optical selection rule $L=0$ means that the emission intensity of
a QR can be dramatically suppressed in $B$-intervals where the
exciton ground state acquires a non-zero momentum.  This
suppression implies that the exciton lifetime becomes
significantly longer, as its dipole-allowed decay is forbidden.
Such behavior is seen in the lower panels of Figs.\ 2 and 3. Thus,
we predict a striking manifestation of the AB effect in the
emission spectra of QR's at low temperatures. Of course, the
emission intensity as a function of the magnetic field strongly
depends on Coulomb correlations in the exciton (compare Figs.\ 2
and 3). However, the appearance of the $B$-windows in the emission
of QR's of various sizes has a common origin -- the electric
dipole moment in the radial direction. The dipole moment is
therefore responsible for the magnetic interference effect while
the exciton moves along the QR. Previous models of QR's with
$R_e=R_h$ could not (and did not) exhibit this important effect
\cite{15}.

Our discussion implies the conservation of angular momentum in a
QR which can be, in principle, destroyed by defects. However, as
mentioned already, recent experiments on optical and electronic
properties of QR's give strong evidence of {\em coherent} motion
of electrons in these self-assembled nanostructures \cite{13}.

It is important to emphasize that these transitions depend on the
net magnetic flux $\Delta \Phi$  through the area {\em between}
the electron and hole trajectories (Fig.\ 1b). Due to the global
conservation of angular momentum, the exciton spinning with
nonzero momentum cannot emit a photon and becomes {\em dark} (very
long lifetime) in well-defined intervals of magnetic field, which
can be observed as noticeable `blinking' PL response for different
fields.  This allows tailoring of excitonic transitions by
switching the ground state from radiative to non-radiative wave
function configurations.

The novel property of the topological Berry phase for a polarized
exciton lies in the complex nature of this quasi-particle. For an
electron and hole in the exciton, the {\em difference} of phases
plays an important role rather than simply the phase of the single
particle. Since the exciton `stores' a photon, such a phase
difference determines whether the QR is bright or dark in optical
emission. This offers the unique opportunity to observe
microscopic electronic phases in optical emission experiments.

\begin{acknowledgments}
We thank Axel Lorke for providing us with the quantum rings AF
micrograph, and P. Petroff, J.P. Kotthaus, A.V. Chaplik, and A.V.
Kalameitsev for useful discussions.  We acknowledge support of
US-DOE grant DE-FG02-91ER45334, and the CMSS Program at OU.
\end{acknowledgments}

\ \

\begin{figure}[b!]
\vspace{8ex}
\includegraphics[width=14cm]{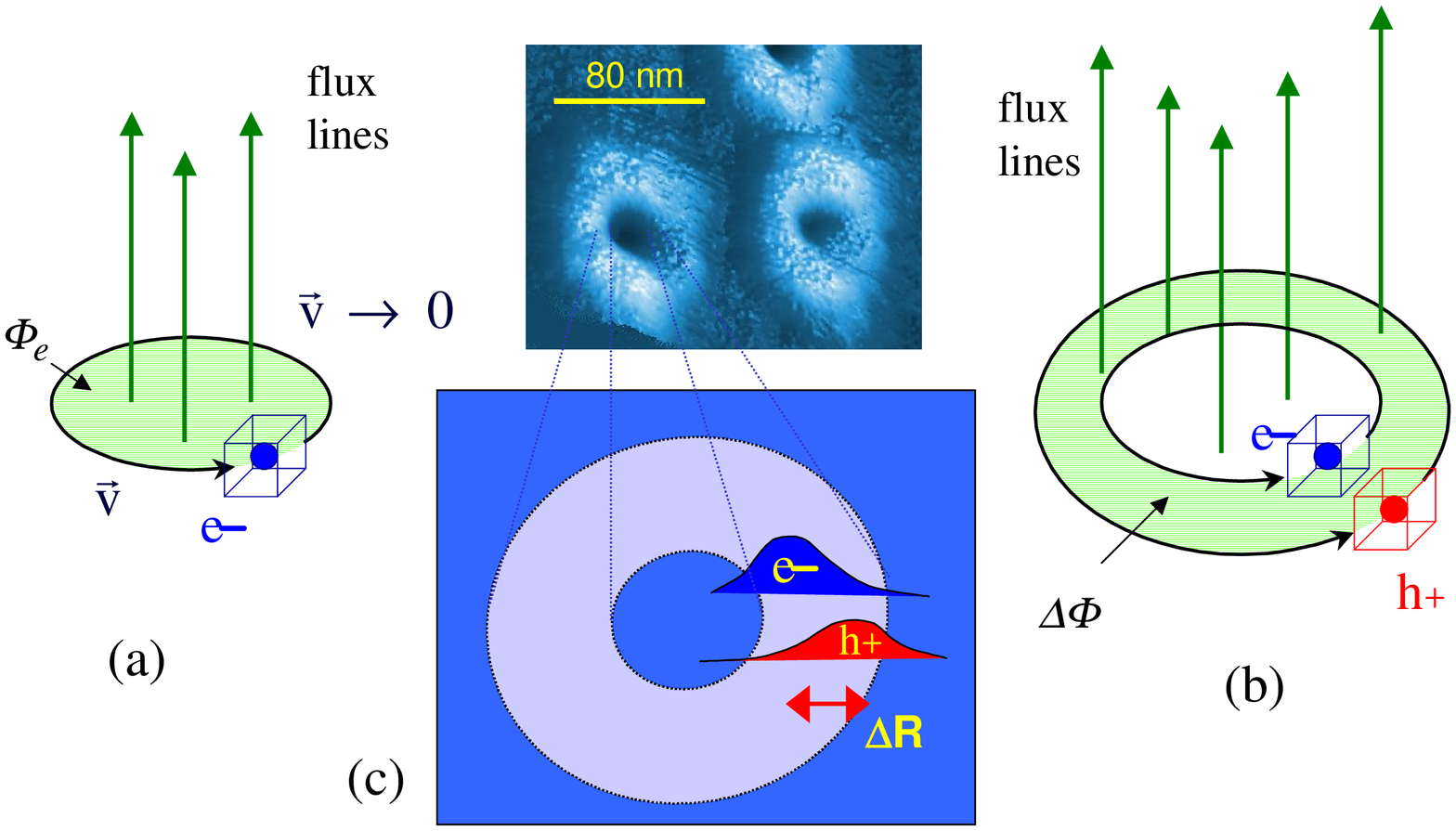}
\vspace{-5ex} \caption{(a) and (b) illustrate origin of Berry
phase for a single charged particle and for an exciton,
respectively. (a) An electron is placed in an isolating box and
rotated along a closed path. The Berry phase, related in this case
to the AB effect, is proportional to the magnetic flux penetrating
the electron trajectory, $\Phi_e$. (b) For the {\em polarized}
exciton, one can adopt the same picture, placing an electron and a
hole in separate boxes. After a full revolution, the electron and
hole accumulate different phases, yielding a \textit{relative}
phase in the pair proportional to the radial dipole moment.  As
the electron and hole induce opposite currents, the AB effect in
this case originates from the magnetic flux through the area
between the two trajectories, $\Delta \Phi$. (c) Atomic-force
micrograph of self-assembled InAs nano-rings on a substrate of
GaAs. Due to different effective masses, the rotating electron and
hole are polarized in the radial direction. The electronic radius
of nanorings is less that the structural one, according to
spectroscopy experiments \cite{13}. }
\end{figure}

\begin{figure}[t!]
\vspace{-5ex}
\includegraphics[width=7cm]{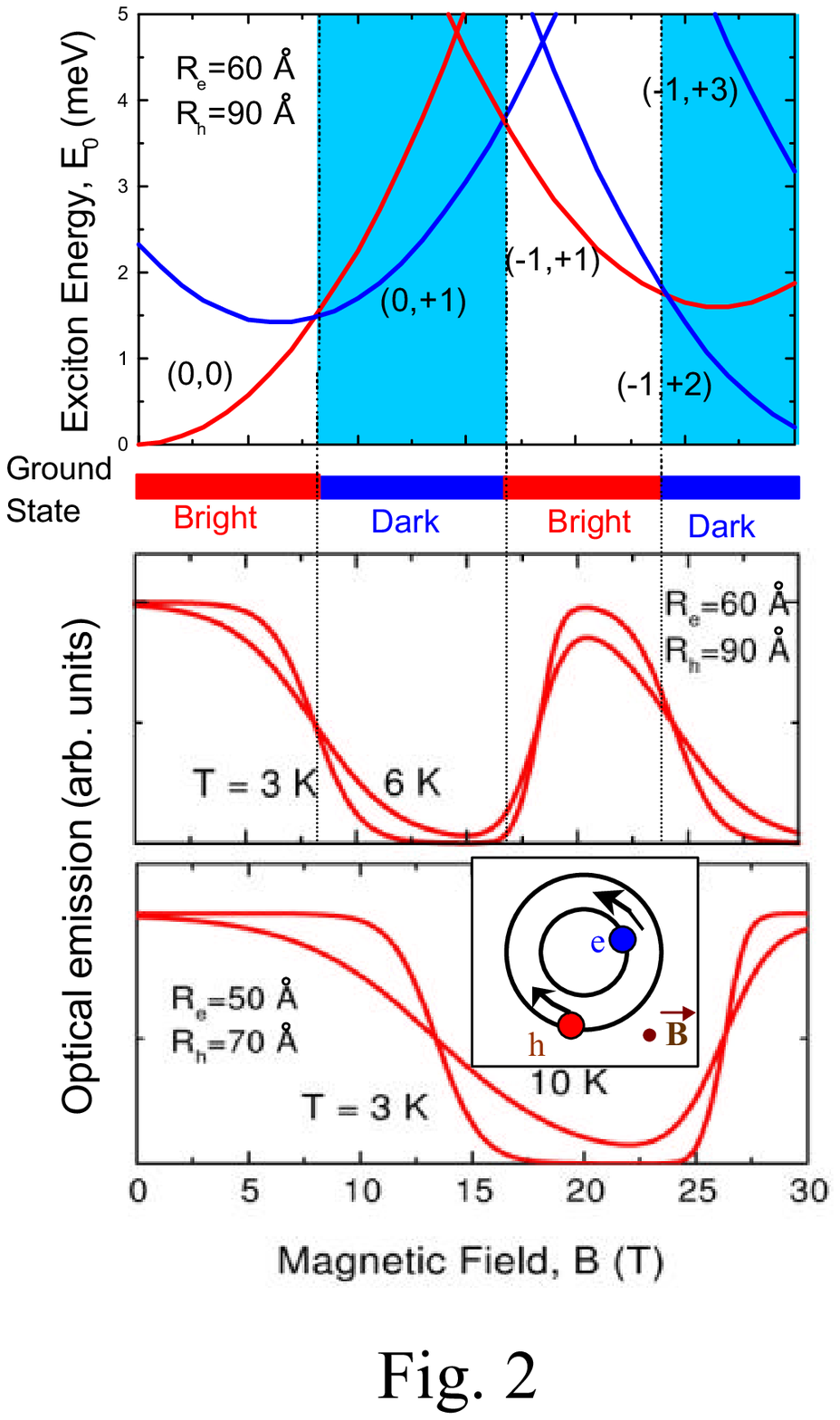}
\vspace{120ex}\caption {Exciton spectrum and emission intensity in
two parallel rings for weak Coulomb interaction (small ring
radii). Upper two panels are for $R_e = 60$\AA ~and $R_h=90$\AA;
lower panel shows emission intensity for a smaller ring. Insert:
uncorrelated motion of particles in a ring. At low temperature $T$
the exciton emission from a ring is strongly suppressed in
well-defined windows of magnetic field which scale with ring
radii. With increasing temperature the excited exciton states
become accessible and the suppression is less pronounced, although
strong oscillations with magnetic field remain. A ``bright" ground
state has $L=0$, and it is ``dark" otherwise.}
\end{figure}

\begin{figure}[t!]
\vspace{-2ex}
\includegraphics[width=9cm]{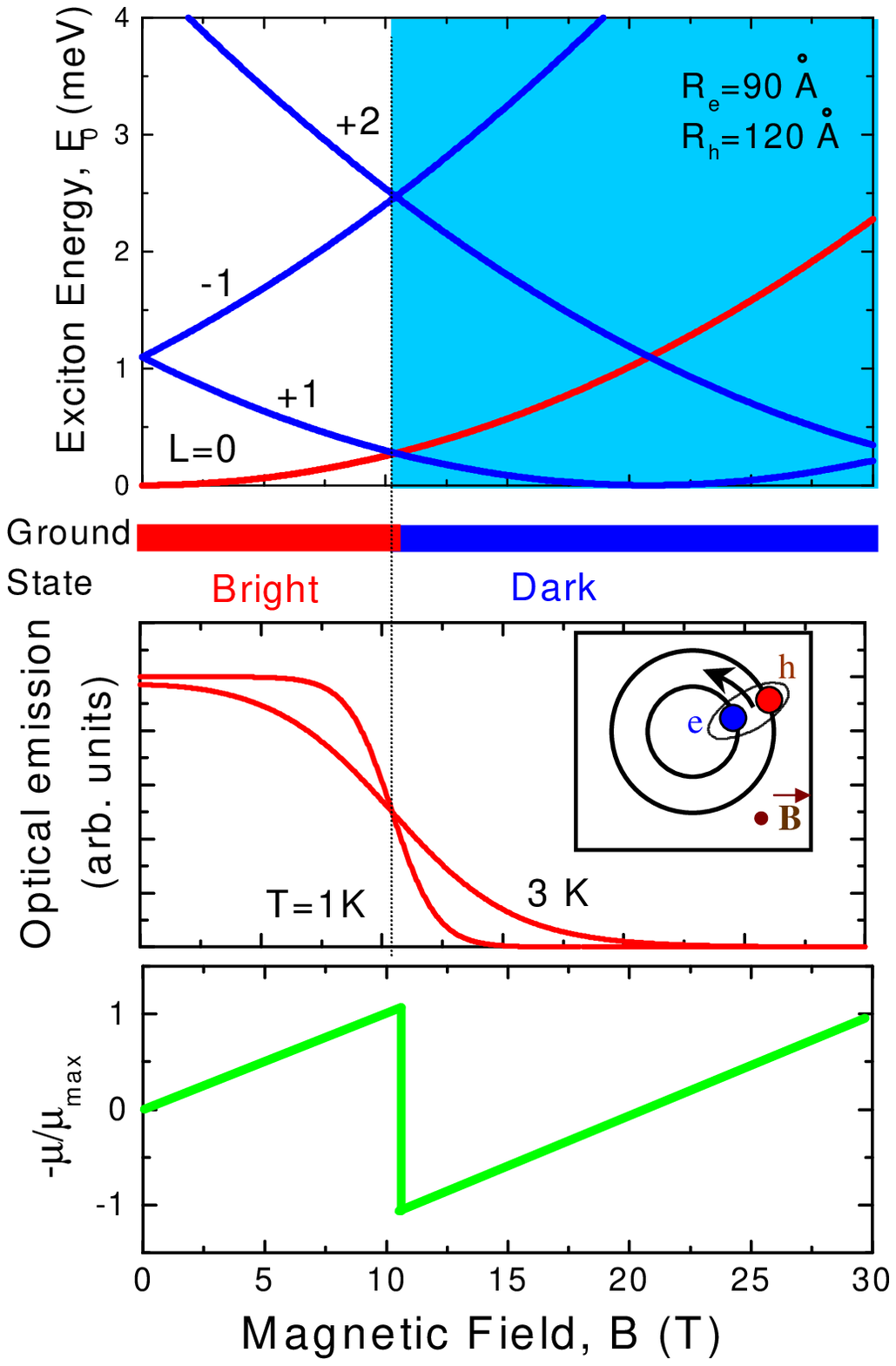}
\vspace{-6ex}\caption{Exciton spectrum, optical emission
intensity, and magnetization in rings for strong Coulomb
interaction (large ring radii). Insert: correlated motion of
electron-hole pair in parallel rings. The character of exciton
spectrum is different to the case of uncorrelated electron-hole
pair (Fig.\ 2). This leads to the existence of a magnetic field
threshold that is well expressed at low temperatures, $T$. Lowest
graph illustrates magnetization of an exciton in the ground state
at low temperature, $\mu = -\partial E_{exc} /
\partial B$, proportional to the persistent current in the
ring \cite{23}. Threshold for emission intensity coincides with
the well-known jump in persistent current vs.\ flux.}
\end{figure}


\vspace{2ex}
\begin{references}
\bibitem{1} See, e.g., J.R. Kirtley, \textit{et al}., Appl. Phys.
Lett. \textbf{66}, 1138 (1995).

\bibitem{2} M.V. Berry, Proc. R. Soc. Lond. \textbf{A392}, 45 (1984).
See also A. Shapere and F. Wilczek, eds., {\em Geometric Phases in
Physics} (World Scientific, Singapore, 1989).

\bibitem{3} Y. Aharonov and J. Anandan, Phys. Rev. Lett. \textbf{58}, 1593
(1987).

\bibitem{4} Y. Aharonov and D. Bohm, Phys. Rev. \textbf{115}, 485 (1959).

\bibitem{5} A. Tonomura, \textit{et al}., Phys. Rev.
Lett. \textbf{56}, 792 (1986).  M. Peshkin and A. Tonomura, {\em
The Aharonov-Bohm Effect} (Springer Verlag, Berlin, 1989).

\bibitem{6} E. Buks, \textit{et al}., Nature \textbf{391}, 871 (1998).

\bibitem{7} D. Loss and E.V. Sukhorukov, Phys. Rev. Lett. \textbf{84}, 1035
(2000).

\bibitem{8} A.W. Holleitner, C.R. Decker, H. Qin, K. Eberl, and R.H.
Blick, Phys. Rev. Lett. \textbf{87}, 256802 (2001).

\bibitem{9} R.J. Warburton, \textit{et al}., Nature \textbf{405}, 926 (2000).

\bibitem{10} M. Bayer, O. Stern, P. Hawrylak, S. Fafard, and A. Forchel,
Nature \textbf{405}, 923 (2000).

\bibitem{11} F. Findeis, \textit{et al}., Appl. Phys. Lett. \textbf{78}, 2958 (2001).

\bibitem{12} D. Mowbray, Phys. World \textbf{13}, 27 (2000);
P. Michler, \textit{et al}., Science \textbf{290}, 2282 (2000).

\bibitem{13} A. Lorke, \textit{et al}., Phys. Rev. Lett.
\textbf{84}, 2223 (2000).

\bibitem{14} C. Obermuller, \textit{et al}., Appl. Phys. Lett.
\textbf{74}, 3200 (1999).

\bibitem{15} A.V. Chaplik, JETP Letters \textbf{62}, 900 (1995); R.A.
R\"omer and M.E. Raikh, Phys. Rev. B \textbf{62}, 7045 (2000).

\bibitem{16} J. Song and S.E. Ulloa, Phys. Rev. B \textbf{63},
125302 (2001); H. Hu, J.L. Zhu, D.J. Li, J.J. Xiong, Phys. Rev. B
\textbf{63}, 195307 (2001).

\bibitem{17} J. Anandan, Phys. Rev. Lett. \textbf{85}, 1354 (2000).

\bibitem{18} J.M. Garcia, \textit{et al}., Appl. Phys. Lett.
\textbf{71}, 2014 (1997).

\bibitem{19} F. Hatami, \textit{et al}., Appl. Phys. Lett.
\textbf{67}, 656 (1995); A.I. Yakimov, {\em et al}., Phys. Rev. B
\textbf{63}, 045312 (2001); M. Hayne, {\em et al}., Phys. Rev. B
\textbf{62}, 10324 (2000).

\bibitem{20} Although we have emphasized intrinsic potential asymmetries
in self-assembled QR's (or type II dots), there are other
alternatives.  The QR can be fabricated by etching technology
which allows the incorporation of top metallic gates. In such
etched systems, a voltage applied to the metallic gate in the
center of a QR may result in a tunable dipole moment for excitons.

\bibitem{21} L. Jacak, P. Hawrylak, and A. Wojs,
\textit{Quantum dots} (Springer Verlag, Berlin, 1998).

\bibitem{22} A.B. Kalameitsev, V.M. Kovalev, and A.O. Govorov, JETP
Lett. \textbf{68}, 669 (1998); K.L. Janssens, B. Partoens, and
F.M. Peeters, Phys. Rev. B \textbf{64}, 155324 (2001).

\bibitem{23} In recent years, persistent currents in mesoscopic loops
have triggered a lot of attention and activity. See e.g., C.
Chapelier, D. Mailly, and A. Benoit, Festkorp.- Adv. Sol. State
Phys. \textbf{34}, 163 (1995).

\end{references}
\end{document}